\documentclass[graybox]{svmult}

\usepackage{type1cm}        % activate if the above 3 fonts are
\usepackage{makeidx}         % allows index generation
\usepackage{graphicx}        % standard LaTeX graphics tool
\usepackage{multicol}        % used for the two-column index
\usepackage[bottom]{footmisc}% places footnotes at page bottom

\usepackage{newtxtext}       % 
\usepackage{newtxmath}       % selects Times Roman as basic font
\usepackage{hyperref}
\usepackage{cite}

\makeindex             % used for the subject index
\renewcommand\thechapter{1}

\begin{document}

\title*{Adversarially Robust Quantum Transfer Learning}
\author{Amena Khatun and Muhammad Usman}

\institute{Amena Khatun \at Quantum Systems, Data61, CSIRO, Australia\\
\email{amena.khatun@data61.csiro.au}
\and Muhammad Usman \at Quantum Systems, Data61, CSIRO, Australia \\
School of Physics, The University of Melbourne, Victoria, Australia\\
\email{muhammad.usman@data61.csiro.au}}

\maketitle
\vspace{-25mm}
Quantum machine learning (QML) has emerged as a promising area of research for enhancing the performance of classical machine learning systems by leveraging quantum computational principles. However, practical deployment of QML remains limited due to current hardware constraints such as limited number of qubits and quantum noise. This chapter introduces a hybrid quantum-classical architecture that combines the advantages of quantum computing with transfer learning techniques to address high-resolution image classification. Specifically, we propose a Quantum Transfer Learning (QTL) model that integrates classical convolutional feature extraction with quantum variational circuits. Through extensive simulations on diverse datasets including Ants \& Bees, CIFAR-10, and Road Sign Detection, we demonstrate that QTL achieves superior classification performance compared to both conventional and quantum models trained without transfer learning. Additionally, we also investigate the model's vulnerability to adversarial attacks and demonstrate that incorporating adversarial training significantly boosts the robustness of QTL, enhancing its potential for deployment in security sensitive applications.
\vspace{-5mm}
\section{Introduction}
\label{sec:1}
Machine learning (ML) has witnessed rapid adoption across numerous domains due to its capability to autonomously learn patterns from data and make predictive decisions. Its transformative impact spans critical areas such as autonomous driving \cite{grigorescu2020survey}, intelligent surveillance systems \cite{sreenu2019intelligent}, facial recognition technologies \cite{schroff2015facenet}, object detection \cite{ren2015faster}, person re-identification \cite{khatun2018deep,khatun2020semantic,khatun2020joint,khatun2021end,khatun2023pose}, behavior analysis \cite{sturman2020deep}, and anomaly detection \cite{pang2021deep}. However, the increasing complexity and size of datasets, combined with the computational intensity of training deep neural networks, have motivated the exploration of Quantum Machine Learning (QML).

\begin{figure*}
\begin{center}
\includegraphics[width=1.0\linewidth]{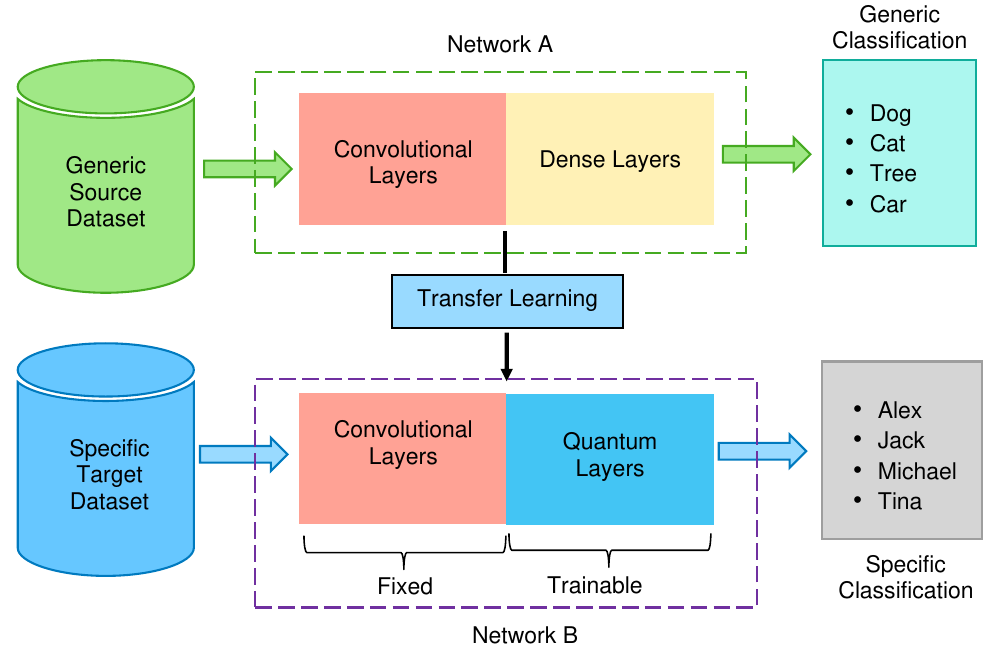}
\end{center}
\caption{Overview of Quantum Transfer Learning (QTL). Network A is trained on a source dataset which is typically large and diverse. The purpose of training Network A is to learn general features, patterns, and representations that are applicable to the broader domain. Here, Network A will learn to recognise basic shapes, textures, and patterns commonly found in images of various animals and objects. Network B is the target model that is used for a different but related task. Instead of starting the training of Network B from scratch, the weights and parameters are initialised and learned by Network A. QTL model performs well on a smaller or completely different dataset compared to the source dataset as it learns features and representations from a source task to initialize a model for the target task. Even if the source and target domains are different, many lower-level features (edges, colors, and textures) and patterns are common to both domains. After initialization, Network B is further trained on the target dataset. During this fine-tuning process, the model adapts its representations to the specific characteristics and requirements of the target task. This fine-tuning allows the model to specialize while retaining the general knowledge from Network A. This figure is reproduced under terms of the CC-BY license. \cite{khatun2025quantum}. Copyright 2024, Amena Khatun and Muhammad Usman. Advanced Quantum Technologies published by Wiley-VCH GmbH.}
\label{fig:QTL}
\end{figure*}

QML employs the principles of quantum mechanics such as superposition, entanglement, and interference to develop fundamentally new learning algorithms \cite{cerezo2022challenges, west2023reflection, ren2022experimental, west2023provably, huang2022quantum, west2023boosted, tsang2023hybrid, khatun2024quantum, wang2024quantum, gill2022quantum, wang2025self}. These algorithms offer the potential for exponential speedups and improved accuracy in solving complex learning problems. A particularly notable development in this space has been the development of hybrid quantum-classical models, including quantum convolutional neural networks (QCNNs), which have demonstrated promising performance in classification tasks while maintaining relatively low model complexity \cite{cong2019quantum, beer2020training}. These models show effectiveness particularly in low-parameter environments \cite{chen2023quantum, hur2022quantum}, yet their scalability to high-dimensional, real-world datasets remains constrained. To overcome these scalability issues and to extend the applicability of QML to more complex datasets, researchers have proposed Quantum Transfer Learning (QTL) as a hybrid strategy that integrtes classical deep learning with quantum neural networks (QNNs) \cite{mari2020transfer, otgonbaatar2022quantum, mogalapalli2022classical, azevedo2022quantum, alex2023hybrid, khatun2025quantum}. In a QTL framework, a classical model often pre-trained on large datasets is used for feature extraction, while a quantum variational circuit handles classification. This effectively reduces the quantum resource burden while still leveraging the advantages of quantum computation.

Transfer Learning (TL) itself has become a widely-used method in deep learning for accelerating training and improving performance, especially when labeled data is limited \cite{bengio2012deep}. It enables the reuse of parameters (such as weights and biases) from a model trained on one task (e.g., Task A using Dataset X) for a related task (e.g., Task B using Dataset Y). By freezing the early layers of a deep network that capture generic visual features like edges, textures, and patterns, and fine-tuning the later layers, TL achieves efficient adaptation to new domains with minimal computational overhead \cite{neyshabur2020being, kornblith2019better}. This strategy significantly reduces training time while improving model generalisation. Figure \ref{fig:QTL} illustrates the overview of TL approach.

Despite the notable performance improvements offered by QML models, handling high-dimensional data remains a key challenge. Firstly, the number of qubits required to represent high-dimensional feature spaces scales with data complexity, but current quantum devices are limited in qubit count. Secondly, increasing the number of input features typically necessitates deeper quantum circuits, which are more susceptible to noise and decoherence. Thirdly, encoding high-dimensional classical data into quantum states efficiently is still an unresolved issue, with ongoing research into effective quantum data encoding strategies \cite{nakaji2022approximate, larose2020robust, wu2023radio, west2023drastic}. These challenges motivate the design of hybrid QTL models that balance quantum computational efficiency with classical preprocessing capabilities. Several QTL approaches have recently been developed with various applications such as waste detection \cite{mogalapalli2022trash}, image-based crack identification \cite{alex2023hybrid}, tuberculosis diagnosis \cite{mogalapalli2022classical}, and breast cancer detection from mammograms \cite{azevedo2022quantum}. Ref. \cite{qi2022classical} presented a QTL model for speech command recognition. While these QTL approaches show promising results, these studies lack a comprehensive comparison between QTL and classical TL methods. Without such comparisons, it is difficult to understand if the observed performance gains in QTL are outperforming the classical approaches. 

Another notable limitation of the current QTL approaches is the lack of research on their vulnerability to adversarial attacks. Adversarial attacks introduce subtle, often imperceptible perturbations to input data to trick the model to make mistakes, which is a serious threat to the reliability of ML and QML systems. Although extensively studied in classical settings \cite{szegedy2013intriguing, huang2011adversarial, DBLP:journals/corr/GoodfellowSS14, DBLP:conf/iclr/KurakinGB17a, ilyas2018black, DBLP:conf/iclr/TjengXT19}, their impact on QML models, especially within the QTL context, remains largely unexplored. This oversight is significant, as such vulnerabilities could undermine the integrity of QML models in critical domains such as defense, autonomous systems, and healthcare.

In light of these gaps, this chapter introduces a comprehensive QTL framework that not only benchmarks the performance of quantum-enhanced TL models against classical baselines but also rigorously examines their susceptibility to adversarial attacks. By employing ResNet-18 as the classical feature extractor and a quantum variational circuit as the classifier, we build a hybrid model that demonstrates superior classification accuracy across datasets such as Ants \& Bees, CIFAR-10, and Road Sign Detection. Furthermore, we evaluate the model's adversarial robustness by subjecting it to adversarial perturbation scenarios and show that incorporating adversarial training (AT) significantly enhances its resilience. This leads to the development of a more robust Quantum Adversarial Transfer Learning (QATL) which learns from adversarial examples during training to mitigate vulnerabilities and ensure more reliable predictions in real-world environments.

\section{Literature Review}
\label{lit_review}

This section provides an overview of related research in hybrid quantum-classical convolutional neural networks, quantum transfer learning, adversarial vulnerability in quantum models, and methods to enhance their robustness through AT.

\subsection{Quantum-Classical Hybrid Convolutional Neural Networks (QCHCNNs)}

The performance of classical ML algorithms is limited by the computational demands of processing large-scale data. Quantum computing, with its potential to perform certain tasks faster than classical systems, offers a promising direction for developing QML models capable of addressing this growing complexity. Hence, the field of QML has attracted substantial attention, with several recent works proposing quantum algorithms aimed at outperforming their classical counterparts \cite{rebentrost2014quantum, biamonte2017quantum, lloyd2014quantum, lloyd2013quantum}. Motivated by these developments, researchers have introduced various quantum convolutional neural network (QCNN) architectures developed to near-term quantum devices. Ref. \cite{cong2019quantum} proposed a QCNN that requires only $O(\log(N))$ variational parameters for $N$ qubits, making it scalable for practical implementations. Their architecture integrates concepts from quantum error correction and multiscale entanglement renormalisation. Similarly, Ref. \cite{kerenidis2019q} developed a quantum clustering algorithm inspired by classical k-means, which can achieve exponential speedups under certain data conditions. Expanding on conventional QCNN architectures, Ref. \cite{maccormack2022branching} introduced the branching QCNN, which enhances feature extraction by leveraging global information from all the qubits that are measured at the pooling layers. This design enables the model to capture long-range correlations between qubits as this approach is not only focusing on the local information. In high-energy physics applications, Ref. \cite{chen2022quantum} proposed a hybrid QCNN for event classification and demonstrated its superior learning efficiency and test accuracy compared to classical CNNs, despite using a similar number of parameters.

Due to the wide range of applications of image classification across domains such as autonomous driving \cite{chen2017multi, maturana2015voxnet}, defense, healthcare \cite{mckinney2020international, esteva2017dermatologist}, and security \cite{taigman2014deepface}, QCNNs have also been developed for image classification tasks. Quanvolutional Neural Network (QNN) is introduced in Ref. \cite{henderson2020quanvolutional}, which incorporates a quantum transformation layer using randomised quantum circuits. This method is evaluated on MNIST dataset, and demonstrated performance improvements in test accuracy and training speed over classical baselines. Further innovations in quantum feature extraction have been proposed by Ref. \cite{chen2023quantum}, that integrates Multi-scale Entanglement Renormalisation Ansatz and fractal feature extraction via box-counting within a QCNN architecture for binary classification. Their model outperformed classical CNNs on a breast cancer dataset. Ref. \cite{hur2022quantum} developed a fully parameterised QCNN using shallow-depth circuits and two-qubit interactions, which is well-suited for noisy intermediate-scale quantum (NISQ) hardware. Their model achieved strong results on MNIST and Fashion-MNIST datasets, despite having a small number of parameters.

\subsection{Quantum Transfer Learning}
While hybrid QCNNs have shown promising capabilities in small-scale image classification tasks, their extension to more complex, real-world datasets is challenging due to current hardware limitations, specifically the small number of available qubits and relatively high noise levels. To address these challenges, in recent years, researchers have explored QTL to combine classical ML scalability with QNNs.

Ref. \cite{mari2020transfer} introduced the concept of QTL and proposed a quantum variational circuit (QVC) to a classical pre-trained model, such as ResNet, to leverage classical feature extraction while performing final classification in the quantum domain. This method proved particularly effective for high-dimensional input data, as the classical component handles heavy pre-processing, thereby minimising the quantum hardware load. Building on this foundation, Ref. \cite{otgonbaatar2022quantum} extended QTL to image classification with small, high-dimensional datasets like Eurosat and UC Merced. They used a hybrid approach involving classical CNNs and multi-qubit QML networks to classify $256\times256\times3$ images, demonstrating improved performance through strong entanglement and amplitude encoding in the quantum layer. Ref. \cite{mogalapalli2022classical} further validated QTL's potentiality by applying it to diverse tasks such as chest X-ray tuberculosis detection, concrete crack recognition, and trash material classification. Their findings underscored that pre-trained classical extractors like DenseNet, AlexNet, and VGG19 provide advantages when combined with a quantum classifier, especially with limited quantum resources. Ref. \cite{azevedo2022quantum} adapted this methodology for breast cancer detection in full-image mammograms, further proving that quantum-enhanced TL can  enhance the performance of purely classical models. Similarly, Ref. \cite{alex2023hybrid} applied a hybrid QTL framework to grayscale crack detection in images and reported improved accuracy, replicating Ref. \cite{mari2020transfer} architecture and showing its generalisability across domains.

Going beyond the visual domains, Ref. \cite{qi2022classical} proposed a QTL architecture for spoken command recognition using a classical 1D CNN for feature extraction and a QVC-based quantum classifier. This work illustrated that QTL's benefits are not limited to image classification but extend to other modalities such as audio processing. Recent advancements have further deepened the understanding of QTL's capabilities. Ref. \cite{zhang2024quantum} proposed a QTL framework using QVC and demonstrated that expressibility linked to circuit depth, structure, and qubit count has a strong correlation with classification accuracy. This study highlights the importance of circuit design in enhancing QTL efficiency and found that annular QVCs combined with a ResNet-18 backbone significantly reduced training time while maintaining high accuracy. Ref. \cite{tseng2025transfer} approached QTL from a theoretical perspective, studying how pre-trained QVCs can be fine-tuned across domains. Ref. \cite{buonaiuto2024quantum} extended QTL into natural language processing by using embedding vectors from large language models to perform sentence acceptability judgments. The QTL model not only outperformed classical performance but also showed better handling of complex sentence structures, pointing toward a quantum advantage in future NLP tasks.

The existing QTL studies have made notable progress in addressing key challenges in image classification, particularly limited qubits and high noise levels. Several QTL approaches have been proposed to leverage the advantages of TL in the quantum field. However, these studies generally do not explore the performance of QTL models in the absence of transfer learning, that is, without the use of pre-trained classical networks. Furthermore, direct comparisons between QTL and their classical counterparts remain limited, hindering a comprehensive understanding of QTL's performance over classical TL. While current QTL methods demonstrate promising potential, more extensive evaluations are needed to assess their practical benefits and limitations in real-world applications. Notably, most existing QTL studies have focused on low-resolution datasets such as EuroSAT ($64\times64\times3$), UC Merced Land Use ($256\times256\times3$), and MNIST ($28\times28\times3$). In contrast, our proposed method evaluates QTL on high-resolution image datasets including Ants \& Bees ($768\times512\times3$) and Road Sign Detection ($1024\times1024\times3$), thereby addressing a significant gap in the literature.

\subsection{Adversarial Vulnerabilities in QML}
It is well established that even the most advanced classical ML models are vulnerable to small, deliberately crafted perturbations applied to even a few pixels of the imput data. These subtle perturbations are referred to as adversarial examples which can lead to significant misclassification, raising serious concerns about the security and robustness of ML applications \cite{szegedy2013intriguing, huang2011adversarial, DBLP:journals/corr/GoodfellowSS14, DBLP:conf/iclr/KurakinGB17a, ilyas2018black, DBLP:conf/iclr/TjengXT19}. This vulnerability raises similar concerns for QML systems, particularly as their adoption grows in sensitive domains. Recent studies have demonstrated that quantum learning systems much like their classical counterparts are susceptible to adversarial manipulation, regardless of whether the input data is classical or quantum in nature \cite{43405, liao2021robust, lu2020quantum, anand2021noise, guan2021robustness, gong2022enhancing, geng2023hybrid, khatun2025classical}. Quantum classifiers, including those that achieve near state-of-the-art performance, can be misled by inputs containing imperceptible perturbations, resulting in incorrect predictions. Ref. \cite{lu2020quantum} systematically investigated the vulnerability of QML models under both white-box and black-box attack scenarios. Their experiments included handwritten digit classification on MNIST, phase recognition in physical systems, and quantum data classification. In the white-box setting, where the attacker has full access to the model architecture and parameters, even minor perturbations led to substantial drops in quantum classifier accuracy. In the black-box setting, adversarial examples generated using classical models were successfully transferred to quantum models, highlighting the transferability of adversarial examples and the inherent susceptibility of quantum classifiers to such attacks. In contrast, Ref. \cite{west2023towards} presented a more optimistic view of QML robustness. By comparing the adversarial performance of QVCs with classical models like CNNs and ResNet-18, their results demonstrated that quantum models exhibit greater resilience to attacks crafted for classical classifiers. This difference was attributed to the distinct feature representations learned by QVCs, which may be less aligned with those exploited by traditional adversarial attacks. These findings suggest that while QML models are not immune to adversarial threats, their internal feature representations may offer some degree of inherent robustness, especially against cross-model attacks.

Despite these advancements, defending QML models from adversarial attacks remains a challenging and open research problem. While theoretical frameworks and empirical results are emerging, practical implementations of adversarial defenses for quantum models are still limited. As QML moves toward real-world applications, ensuring the robustness and security will be critical for their deployment in security-sensitive applications.

\subsection{Quantum Adversarial Training}
Enhancing the robustness of quantum classifiers is essential to ensure the security and reliability of QML applications. While AT has been widely recognised as an effective defense mechanism in classical ML, its adoption in the quantum domain remains limited. Only a few of studies have explored its applicability and impact within QML frameworks \cite{lu2020quantum, ren2022experimental, west2023benchmarking}. Ref. \cite{lu2020quantum} demonstrated the vulnerability of quantum classifiers to adversarial perturbations, comparable to their classical counterparts. The study first introduced AT in QML and reported a notable improvement in  robustness of quantum classifier against adversarial attacks generated using the Basic Iterative Method \cite{feinman2017detecting}. However, the evaluation was limited to binary classification tasks on the downscaled MNIST dataset and did not explore the model's applicability to more complex, multi-class, and high-dimensional image datasets. Expanding on this foundation, Ref. \cite{west2023benchmarking} presented a comprehensive framework to assess and benchmark the adversarial robustness of both quantum and classical ML models. This study evaluated QVCs, CNNs, and ResNet-18 architectures under both classical and quantum adversarial attacks, including Fast Gradient Sign Method (FGSM) \cite{43405}, Projected Gradient Descent \cite{madry2017towards}, and AutoAttack \cite{croce2020reliable}. The results showed that while AT significantly enhanced the robustness of classical models against classical attacks, the improvements observed in QVCs were relatively modest. The authors concluded that QML models, particularly QVCs, exhibit inherent robustness against certain types of adversarial perturbations, and the additional benefits provided by AT are limited in such cases. These findings suggest different interpretations, while Ref. \cite{lu2020quantum} reported substantial gains from AT, especially under BIM attacks in a binary setting, Ref. \cite{west2023benchmarking} found that quantum models are already resilient to many classical attacks and therefore do not benefit substantially from AT in more diverse and realistic settings. Notably, the methodological scope of these two works differs significantly in terms of dataset complexity, number of classes, and variety of attack methods. Although early research supports the use of AT as a defense mechanism in QML, its overall effectiveness across architectures and datasets remains an open research question. A systematic investigation involving diverse QML architectures, multi-class datasets, and a range of adversarial attack models is necessary to fully understand the impact and limitations of AT in the quantum setting.

\section{Methodology} 
\label{method}

\subsection{Knowledge Transfer from Classical CNNs to QCHCNNs}
\label{method-a}
\begin{figure*} 
\begin{center} 
\includegraphics[width=1.0\linewidth]{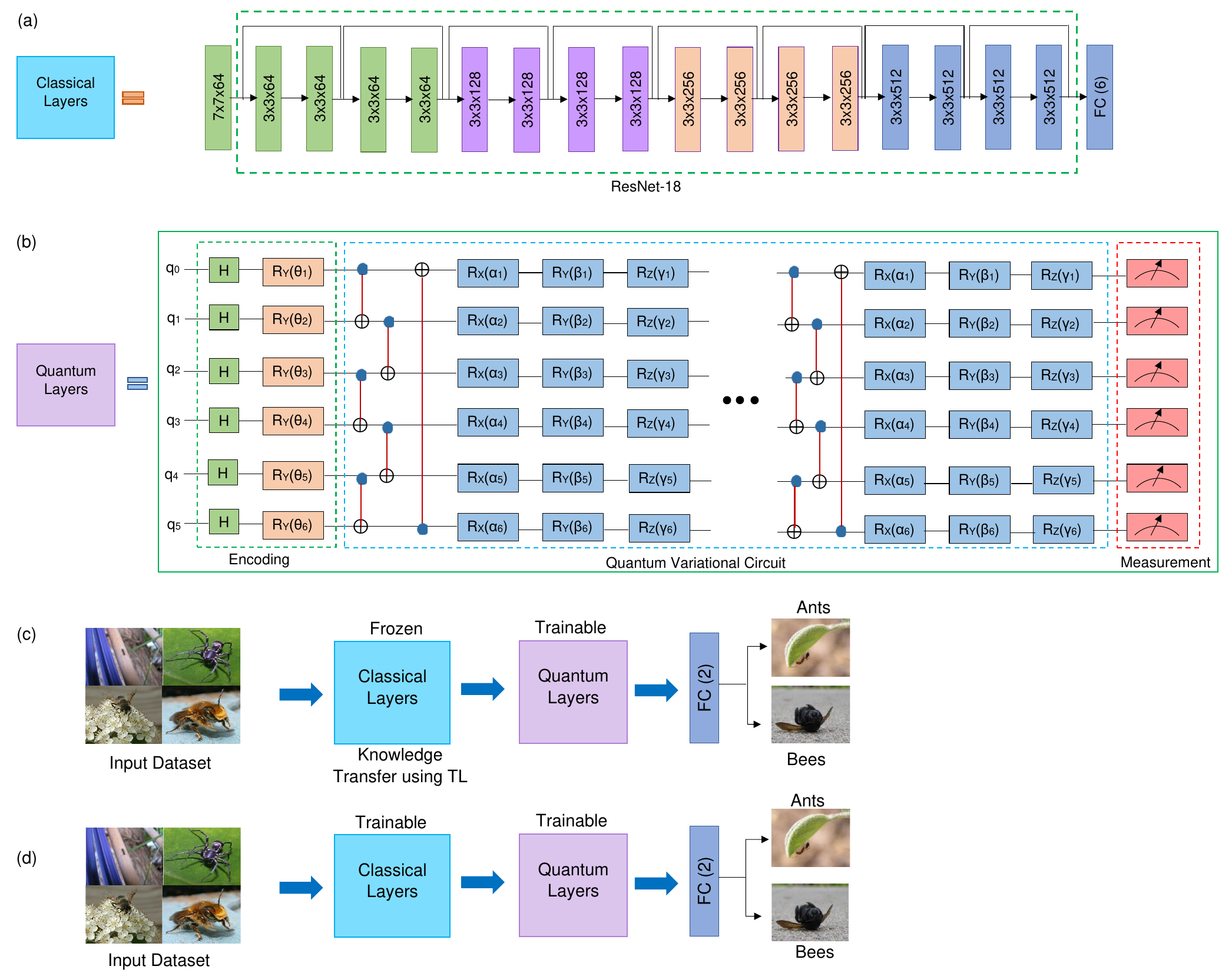} 
\end{center} 
\caption{The architecture of the proposed QTL framework. (a) illustrates the classical layers utilizing ResNet-18, which comprises 18 layers organized in residual blocks with skip connections. (b) depicts the trainable quantum components, including data encoding, a Quantum Variational Circuit (QVC), and measurement layers. The QVC integrates CNOT gates and single-qubit rotation gates $(R_X, R_Y, R_Z)$. (c) shows the complete QTL workflow, where input data passes through frozen ResNet-18 layers before reaching trainable quantum layers for final classification. (d) demonstrates the architecture of the hybrid quantum-classical model without transfer learning, where all classical and quantum layers are updated during training. The Ants and Bees dataset used is publicly available at \url{https://www.kaggle.com/datasets/gauravduttakiit/ants-bees}. This figure is reproduced under terms of the CC-BY license. \cite{khatun2025quantum}. Copyright 2024, Amena Khatun and Muhammad Usman. Advanced Quantum Technologies published by Wiley-VCH GmbH.} 
\label{fig:QTL_Architecture} 
\end{figure*}
The proposed QTL framework incorporates TL principles to train QCHCNNs, effectively leveraging the representational power of large datasets to improve classification on specific target datasets such as Ants \& Bees, Road Sign Detection, and CIFAR-10. In this architecture, a widely used deep residual network, ResNet-18 \cite{he2016deep} is employed as a pre-trained model to transfer knowledge from a source domain to the target domain. Residual connections in ResNet significantly improve gradient flow through the network, ensuring stable learning and faster convergence. These connections allow information to bypass certain layers, minimising training time and offer optimal parameter tuning. While deeper networks typically suffer from degradation in performance as their depth increases, ResNet mitigates this through the use of residual mappings, enabling the network to model complex relationships without performance saturation \cite{he2016deep}. The ResNet-18 model is pre-trained on the extensive ImageNet dataset \cite{russakovsky2015imagenet}, which contains over 1.2 million training images across 1,000 classes, and serves as a highly capable feature extractor \cite{neyshabur2020being, kornblith2019better, huh2016makes}. The initial processing of input data involves a convolutional layer with a $7\times7$ filter and stride of 2, followed by a max-pooling layer with a $3\times3$ filter and stride of 2, as shown in Figure \ref{fig:QTL_Architecture}(a). Each subsequent layer contains two residual blocks, each with two weighted layers connected through skip connections and ReLU activations. When input and output dimensions match, identity shortcuts are used. In the case when dimensionality is increased, a
projection shortcut technique is employed by  $1\times1$ convolutions to match the dimension. Downsampling is performed via convolutional layers with a stride of 2, and batch normalisation follows each convolutional layer. The core residual block operation is mathematically represented by:
\begin{equation}
y=\sum F(x,{W_i})+x,
\label{eq:1}
\end{equation} 
where $x$ denotes the input, $y$ the output, and $F(x, W_i)$ is the residual function parameterised by weights $W_i$. The final output from ResNet-18 is a 512-dimensional feature map, which is flattened and reduced to a six dimensional feature vector using a fully connected layer. These six features are then passed into the quantum variational circuit for the classification task, as illustrated in Figure \ref{fig:QTL_Architecture}(c). During training, all ResNet-18 layers remain frozen except for the last dense layer, enabling the network to function purely as a feature extractor and reducing computational costs. As the pre-trained ResNet-18 model is already trained on a dataset with a lot of diverse image categories, the pre-train network will act as a good feature extractor for our data (Ants \& Bees, Road Sign Detection, CIFAR-10), even if the new data are from completely different classes.

\vspace{-4mm} 
\subsection{Quantum-Classical Hybrid Convolutional Neural Network (QCHCNN)} 

The proposed quantum framework consists of three standard stages: classical data encoding into quantum states, quantum variational circuit, and quantum measurement. Angle encoding is used to map classical features into quantum states. With six extracted features, six qubits are initialised, each associated with a corresponding feature. To initiate the quantum states, Hadamard gates are applied to bring qubits into superposition states, followed by $R_Y$ rotation gates to encode the classical feature values as rotation angles.

The encoded quantum state is expressed as:
\begin{align}
|\theta\rangle=|\theta_1\rangle \otimes |\theta_2\rangle \otimes |\theta_3\rangle \otimes |\theta_4\rangle\otimes |\theta_5\rangle\otimes |\theta_6\rangle \nonumber  \\
= \begin{bmatrix}
cos(\theta_1) \\ sin(\theta_1) \\
\end{bmatrix}  \otimes \begin{bmatrix}
cos(\theta_2) \\ sin(\theta_2) \\ 
\end{bmatrix} \otimes 
\begin{bmatrix}
cos(\theta_3) \\ sin(\theta_3) \\
\end{bmatrix}   \otimes 
\begin{bmatrix}
cos(\theta_4) \\ sin(\theta_4) \\
\end{bmatrix} \otimes 
\begin{bmatrix}
cos(\theta_5) \\ sin(\theta_5) \\
\end{bmatrix}  \otimes 
\begin{bmatrix}
cos(\theta_6) \\ sin(\theta_6) \\
\end{bmatrix} \nonumber  \\
= (\otimes_{i=1}^6 R_y(2x_i))|\theta_1\rangle^{\otimes^6}.
\label{eq:2}
\end{align}

Thus, the final encoded quantum state becomes:
\begin{align}
(\otimes_{i=1}^6 R_y(2x_i))|\theta_1\rangle^{\otimes^6}
= \begin{bmatrix}
cos(\pi\theta_1) \\ sin(\pi\theta_1) \\
\end{bmatrix}  \otimes \begin{bmatrix}
cos(\pi\theta_2) \\ sin(\pi\theta_2) \\ 
\end{bmatrix} \otimes 
\begin{bmatrix}
cos(\pi\theta_3) \\ sin(\pi\theta_3) \\
\end{bmatrix}  \otimes 
\begin{bmatrix}
cos(\pi\theta_4) \\ sin(\pi\theta_4) \\
\end{bmatrix}  \otimes 
\nonumber  \\
\begin{bmatrix}
cos(\pi\theta_5) \\ sin(\pi\theta_5) \\
\end{bmatrix}  \otimes 
\begin{bmatrix}
cos(\pi\theta_6) \\ sin(\pi\theta_6) \\
\end{bmatrix}.
\label{eq:2}
\end{align}

A QVC formulates the optimisation problem directly within the quantum computational framework by embedding tunable gate parameters into the circuit design. These parameters are subsequently refined through classical optimisation techniques to identify optimal solutions for the targeted task. The QVC employs a hybrid structure of two types of quantum operations: entangling gates and single-qubit rotation gates. Specifically, the Controlled-NOT (CNOT) gate creates entanglement between pairs of qubits, enabling quantum correlations that surpass the capabilities of classical systems. Here, the single-qubit rotation gates are $R_X$, $R_Y$, and $R_Z$ which apply rotations around the X, Y, and Z axes, respectively. The rotation angles, $\alpha_i$, $\beta_i$, and $\gamma_i$ control these operations and are treated as trainable parameters throughout the learning process. By fine-tuning these angles, the state of each qubit can be manipulated to effectively encode complex quantum information.

The output quantum states from the QVC circuit are evaluated by computing the expectation values of six observables, denoted as $z=[z_1,z_2,z_3,z_4z_5,z_6]$. These observables are selected to capture distinct characteristics of the quantum system. The objective is to estimate the average values of these observables, representing the expected outcomes obtained from repeated measurements of the same quantum state.
\begin{equation} 
M: |\theta\rangle \rightarrow z = \langle \theta | z | \theta \rangle, 
\end{equation} 
where $M$ represents the measurement operation. The QNN model is optimised using stochastic gradient descent (SGD), updating parameters according to:
\begin{equation} 
\theta(t+1) = \theta(t) - \eta \cdot \nabla L(\theta(t)), 
\label{eq:5} 
\end{equation} 
where $\theta(t)$ is the vector of model parameters at iteration $t$, $\theta(t+1)$ is the updated vector of model parameters at iteration $(t+1)$, $\eta$ is the learning rate, a hyperparameter that determines the step size for the update, $\nabla L(\theta(t))$ is the gradient of the loss function $L$ with respect to the model parameters. For a small value of $\epsilon$, the finite difference method is used to approximate the partial derivative as,
\begin{align}
\frac{\partial L(\theta)}{\partial \theta} \approx \frac{L(\theta + \varepsilon) - L(\theta)}{\varepsilon}
\label{eq:2}
\end{align}
Finally, the cross-entropy loss, a commonly used objective function in classical ML, is employed to quantify the discrepancy between the predicted probability distribution and the actual distribution of the target classes.
For binary classification, the loss function is defined as:
\begin{equation} 
L_{classification} = -[y \cdot \log(p) + (1 - y) \cdot \log(1 - p)]
\label{eq:7} 
\end{equation} 
Here, $y$ represents the true class label, $p$ denotes the predicted probability of the sample belonging to the positive class (i.e., class 1), and $1-p$ corresponds to the predicted probability of the sample belonging to the negative class (i.e., class 0). When $y = 1$, the loss depends solely on $\log(p)$, encouraging the model to predict $p$ close to 1, indicating a correct positive classification. Conversely, when $y = 0$, the loss is determined by $-\log(1 - p)$, pushing the model to predict $p$ near 0, which reflects a correct negative classification.
For multi-class scenarios, the loss can be represented as,
\begin{equation} 
L_{classification} = - \sum_{i=1}^{n} p_i \log \bar{p_i} = - \log \bar{p}_t, 
\end{equation} 
where $p_i$ is the true distribution, $\bar{p_i}$ is the predicted probability for class $i$, and $t$ is the target class.

\subsection{QCHCNN without Transfer Learning}
\label{without_TL}

To conduct a comprehensive comparison with the proposed QTL framework, we also design a hybrid classical-quantum architecture that operates without incorporating transfer learning. In this configuration, although ResNet-18 serves as the initial feature extractor, the entire model, including all layers of ResNet-18, is fine-tuned from scratch rather than leveraging pre-trained weights. As illustrated in Figure \ref{fig:QTL_Architecture}(d), all network parameters, both classical and quantum, are updated throughout the training process. The classical component of this hybrid architecture performs the initial preprocessing of raw input data and encodes it into a representation suitable for quantum processing. While ResNet-18 generates high-dimensional feature maps that capture complex hierarchical patterns, directly working with such high-dimensional data can be computationally prohibitive and inefficient. To address this, the extracted features are compressed into a lower-dimensional latent space using a fully connected (Dense) layer, which learns meaningful and compact representations. These compressed features are subsequently passed to the quantum module, where quantum encoding techniques prepare the data for quantum state manipulation. After processing within the quantum circuit, measurement operations are performed to extract classical outputs from the resulting quantum states. These outputs are then forwarded to a final classical fully connected layer, connected through a non-trainable matrix, which integrates quantum-derived features and produces the final predictions. The use of a non-trainable matrix ensures that quantum information is seamlessly incorporated without introducing additional learnable parameters at this stage. Training this hybrid classical-quantum architecture involves jointly optimising both the classical layers and the quantum circuit parameters. The learning process is driven by a loss function that determines the difference between predicted outputs and ground truth labels. Standard gradient-based optimisation algorithms are employed to iteratively adjust the network parameters, minimising the loss and improving predictive accuracy.

However, due to the complexity of quantum state manipulation and the high-dimensional optimisation landscape, training quantum circuits presents significant computational challenges. Techniques such as the parameter-shift rule and classical gradient descent methods are utilised to fine-tune the parameters of the quantum circuit. Unlike the QTL framework, which benefits from faster convergence and improved performance through TL, this architecture demands greater computational resources and more extensive training iterations as it learns classical and quantum representations entirely from the ground up.

\subsection{\textcolor{black}{Classical Transfer Learning}}
\label{classicla_TL}

\begin{figure*}
\begin{center}
\includegraphics[width=0.8\linewidth]{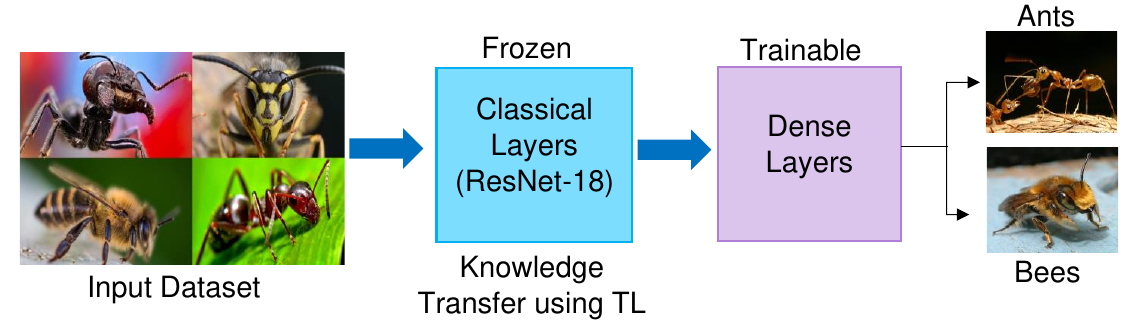}
\end{center}
   \caption{\textcolor{black}{The architecture of classical transfer learning. ResNet-18 consists of 18 layers as illustrated in Figure \ref{fig:QTL_Architecture}(a) and pre-trained on a larger and diverse ImageNet dataset. The knowledge of the pre-trained network is transferred from source domain to the target domain. For classical TL, all the layers of ResNet-18 are frozen except the last dense layer. Only the dense layers are trainable which are fully connected layers for the final classification task. When there is no transfer learning involved, all the classical layers and dense layers are trainable. This figure is reproduced under terms of the CC-BY license. \cite{khatun2025quantum}. Copyright 2024, Amena Khatun and Muhammad Usman. Advanced Quantum Technologies published by Wiley-VCH GmbH.}}
\label{fig:classicalTL}
\end{figure*}

The classical transfer learning architecture is illustrated in Figure \ref{fig:classicalTL}. This approach utilises ResNet-18 as the backbone of the model, pre-trained on the extensive and diverse ImageNet dataset, which contains millions of labeled images spanning a wide array of object categories. In this framework, the pre-acquired knowledge from the source domain is effectively leveraged to improve performance on a new, yet related, target domain task. The detailed design and functionality of ResNet-18 have already been discussed comprehensively in Section \ref{method-a}. During the fine-tuning process, all convolutional layers of ResNet-18 remain frozen, with only the final dense layers left trainable. These fully connected layers act as the new classifier, specifically adapted for the target domain task. In contrast, when TL is not applied, the entire ResNet-18 model, including all convolutional and dense layers, is trained end-to-end from scratch.

\subsection{ Adversarial Attacks on Quantum Models}
\label{adversarial_attack}

\begin{figure*}
\begin{center}
\includegraphics[width=1.0\linewidth]{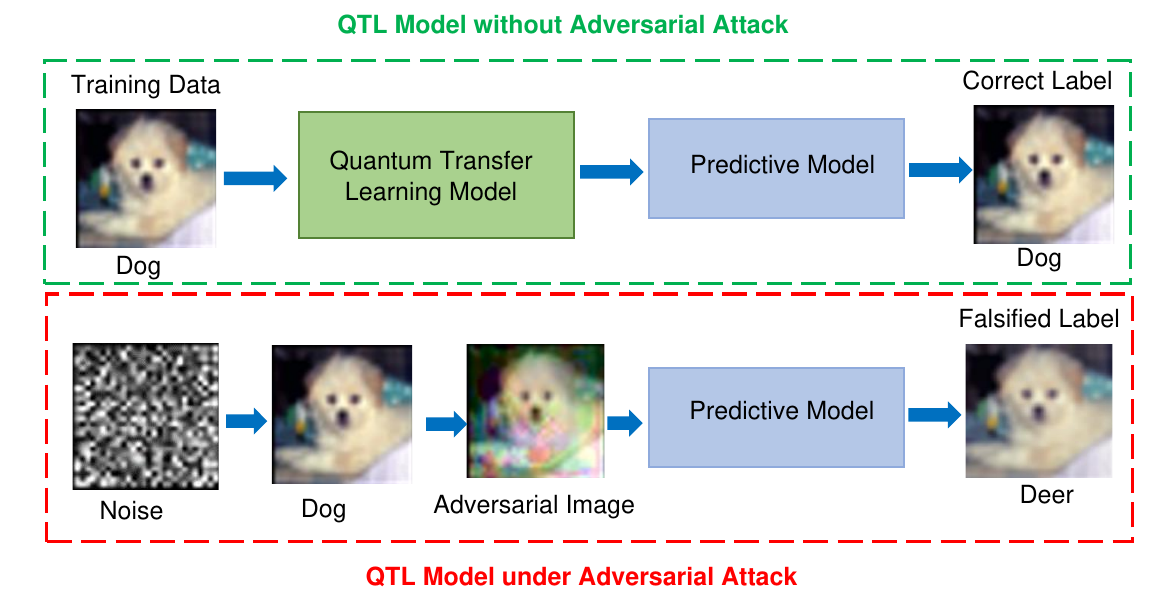}
\end{center}
   \caption{Overview of quantum adversarial attack. If we consider the input image of a dog, the QTL model correctly classifies it as a dog when the model is not under adversarial attack. However, a carefully crafted perturbation is applied to the input image in a way that exploits the vulnerabilities of the proposed QTL model. When the perturbed images are fed to the model for prediction, the QTL model incorrectly classified the image as a deer rather than a dog. This is an indication that the QTL model is vulnerable to small changes in input data, leading to significant implications for the reliability and security of the model. The original image of dog in this figure is from the  publicly available dataset, CIFAR-10 \cite{krizhevsky2009learning}. This figure is reproduced under terms of the CC-BY license. \cite{khatun2025quantum}. Copyright 2024, Amena Khatun and Muhammad Usman. Advanced Quantum Technologies published by Wiley-VCH GmbH.}
\label{fig:adversarial_attack}
\end{figure*}

Adversarial attacks are intentionally crafted to deceive ML models, particularly the classifiers, by introducing subtle perturbations to input data that lead to incorrect predictions. Consider a classifier defined as  $f:X \rightarrow Y$, where $X$ is a subset of $R^d$, representing the input space, and $Y$ denotes the set of possible class labels. The goal of an adversarial attack is to construct a perturbed input $x^\wedge$ that remains close to a valid input sample $x$ under a specified distance metric, while ensuring that the classifier's output for the perturbed input differs from its original prediction, i.e., $f(x^\wedge) \not = f(x)$. If the attacker's sole objective is to induce any incorrect classification without targeting a specific label, it is classified as a non-targeted attack. Conversely, in a targeted attack, the adversary aims to manipulate the classifier so that the perturbed input $x^\wedge$ is misclassified as a predetermined target label $y$, can be represented as $f(x^\wedge) = y$. Regardless of the attack type, the perturbation introduced to the original input should remain minimal to avoid detection. This perturbation is commonly measured using the $L_p$ norm, where typical values of $p$ include 0, 1, 2, or $\infty$. These norms characterise the perturbation in different ways: the $L_1$ norm measures the sum of absolute differences, the $L_2$ norm corresponds to the Euclidean distance, and the $L_\infty$ norm measure the maximum absolute difference across all dimensions. An illustration of adversarial attack process is provided in Figure \ref{fig:adversarial_attack}.
In our proposed QTL framework, we utilise the FGSM attack to generate adversarial examples. FGSM introduces perturbations by modifying the input data in the direction that most rapidly increases the loss of the model, thereby maximising the likelihood of misclassification. Given a classifier $f:X \rightarrow Y$, FGSM computes the gradient of the loss function $L$ with respect to the input sample $x$ as follows:
\begin{align}
\Delta_x L(y, f(x; \theta)),
\label{eq:9}
\end{align}
where $y$ denotes the true class label and $\theta$ represents the model parameters. This gradient identifies the direction in which small changes to the input will most significantly affect the loss.The adversarial example  $x^\wedge$ is then generated by applying a perturbation in the direction of the gradient sign:
\begin{align}
x^\wedge = x + \epsilon \cdot \text{sign}(\nabla_x L(y, f(x; \theta))),
\label{eq:9}
\end{align}
where $\epsilon$ is a small positive constant that controls the magnitude of the perturbation. The $sign$ function extracts the direction of the gradient for each input feature, indicating whether the perturbation should increase or decrease the corresponding input value. This process effectively leads the model to make incorrect predictions while maintaining the perturbed input visually close to the original data.

\subsection{Adversarial Training for Quantum Robustness}
\label{adversarial_training}

\begin{figure*}
\begin{center}
\includegraphics[width=0.98\linewidth]{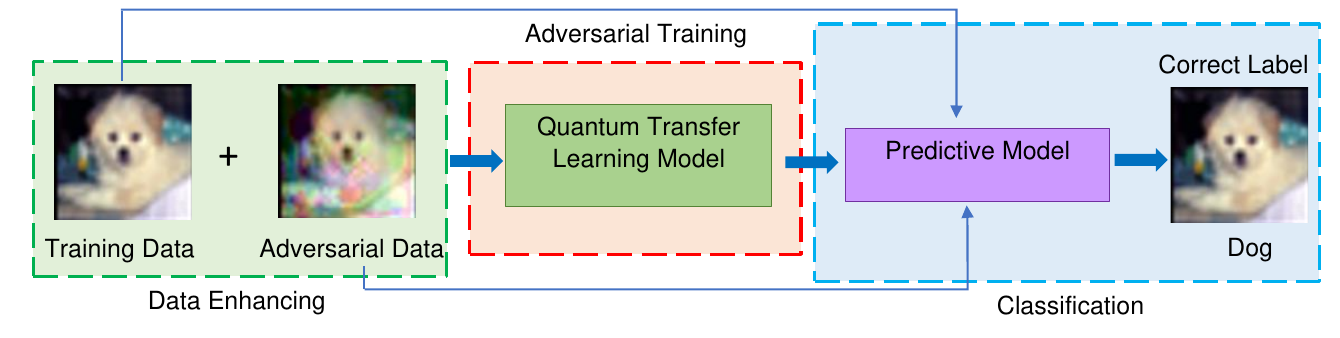}
\end{center}
   \caption{Illustration of quantum adversarial transfer learning architecture. For adversarial training, the training data now consists of both original input images and adversarial examples. As the QTL network is trained on this combined dataset, the network encounters both original and adversarial images. Thus, the network learns to resist the effects of adversarial perturbations and make accurate predictions even when presented with adversarial examples. As a result of adversarial training, the QTL network becomes more robust against adversarial attacks. When presented with a perturbed image that was crafted to deceive the network, it is now less likely to be misclassified. In the figure, where a dog image was misclassified as a deer, the adversarial training would help the network correctly classify it as a dog. The original image of dog in this figure is taken from the  publicly available dataset, CIFAR-10 \cite{krizhevsky2009learning}. This figure is reproduced under terms of the CC-BY license. \cite{khatun2025quantum}. Copyright 2024, Amena Khatun and Muhammad Usman. Advanced Quantum Technologies published by Wiley-VCH GmbH.}
\label{fig:adversarial_training}
\end{figure*}

Adversarial training is a widely adopted defense mechanism in ML aimed at improving the resilience of neural networks against adversarial perturbations. These perturbations introduce subtle, carefully crafted modifications to input data with the intent of misleading a model's predictions. The conceptual framework of adversarial training is illustrated in Figure \ref{fig:adversarial_training}, where the proposed hybrid QTL model is trained using a combination of both original and adversarially perturbed images. Here, we consider a scenario in which the QTL network is tasked with classifying animal images. An adversarial attack might introduce imperceptible changes to an image of a dog, causing the model to incorrectly classify it as a deer. The purpose of adversarial training is to improve the model's robustness against such deceptive manipulations, ensuring reliable predictions even in the presence of adversarially modified inputs. This is achieved by exposing the model to adversarial examples during training, allowing it to learn representations that are less sensitive to small input perturbations.
Mathematically, adversarial training formulates the learning objective as a min-max optimisation problem:
\begin{align}
\min_{\theta} \max_{\delta \in \Delta} \mathcal{L}(f_{\theta}(x + \delta)), y,
\label{eq:11}
\end{align}
where $\theta$ represents the model parameters, $x$ is the original input sample, $y$ is the corresponding ground truth label, and $f_{\theta}(x)$ denotes the model's prediction. The loss function $\mathcal{L}$ calculates the error between the model's predictions and the true labels. The term $\delta$ represents the perturbation applied to the input, and $\Delta$ defines the set of allowable perturbations under the threat model. In this context, the threat model is specified as $\Delta = {\delta : |\delta|{\infty} \leq \varepsilon}$, constraining the perturbations to have a maximum magnitude bounded by $\epsilon$ under the $\mathcal{L}{\infty}$ norm. Directly solving the inner maximisation over the perturbation space $\Delta$ is computationally intensive. Therefore, FGSM is used as an efficient approximation to generate adversarial perturbations. The optimal perturbation is computed as:

\begin{align}
\delta^* = \varepsilon \cdot \text{sign}(\nabla_x \mathcal{L}(f(x), y)),
\label{eq:9}
\end{align}
where $\nabla_x$ denotes the gradient of the loss function with respect to the input $x$. This gradient identifies the direction in which the input should be perturbed to maximise the loss. During training, the model parameters $\theta$ are updated using gradient descent, taking into account the adversarially perturbed inputs generated by $\delta^*$. After each update, the perturbation $\delta$ may push the modified input outside the permissible bounds defined by $\Delta$. To ensure the perturbation remains within the threat model, it is projected back into the allowed set by clipping each element of $\delta$ to the range $[-\varepsilon, \varepsilon]$. By iteratively applying this process, adversarial training strengthens the model’s defenses against adversarial attacks, enabling it to maintain reliable classification performance even when exposed to adversarially manipulated inputs.

\section{Network Setup and Training}

Given the current limitations of quantum hardware such as the limited number of qubits, high error rates, and decoherence, all experimental simulations were performed using PennyLane \cite{bergholm2018pennylane}, an open-source QML framework developed by Xanadu. The models were trained for a total of 50 epochs, requiring approximately 12 minutes and 27 seconds to complete. For the classical components of the framework, we used PyTorch \cite{paszke2019pytorch}.The Adam optimiser was employed across all experiments, using a batch size of 16 and an initial learning rate of 0.0004. This learning rate was kept constant during the first 25 epochs and then gradually decayed linearly to zero over the remaining 25 epochs to ensure smooth convergence. In the quantum circuit configuration, we initialised the model with 6 qubits and a circuit depth of 6 variational layers. Consequently, the total number of trainable parameters for the QVC is 108, calculated as $6 \times 6 \times 3$. In comparison, the classical TL baseline included 1,026 trainable parameters. While increasing the depth of the quantum circuit could potentially improve performance, it also increases the risk of encountering the well-known barren plateau problem, where the optimisation landscape becomes flat, leading to ineffective training. However, since the QVC models demonstrated strong performance during both training and evaluation, we believe further increasing the circuit depth would likely yield only marginal gains. To further support efficient convergence, a learning rate scheduling strategy was applied, reducing the learning rate by a factor of 0.1 every 10 epochs. The quantum model weights were initialised with a small random spread, set to 0.01, to ensure stable optimisation at the start of training.

\begin{figure*}
\begin{center}
\includegraphics[width=1.0\linewidth]{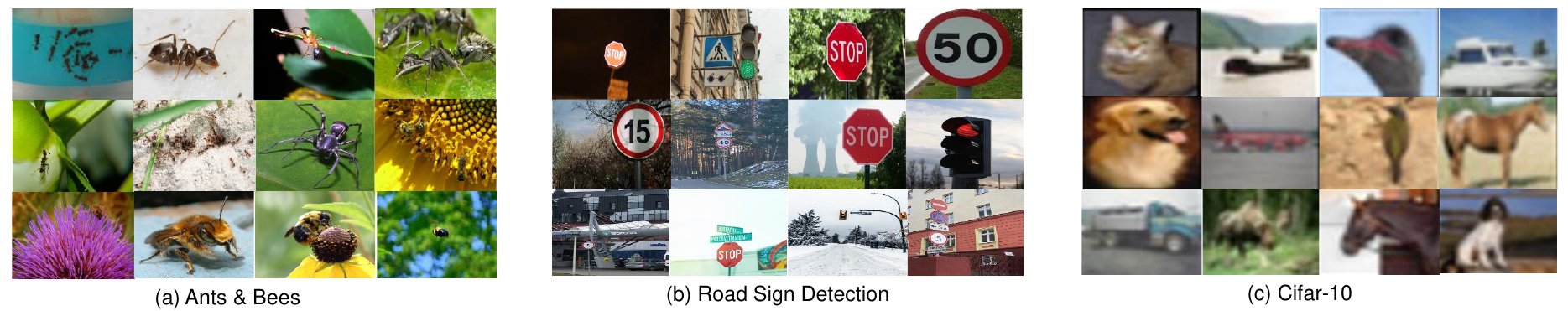}
\end{center}
   \caption{Overview of datasets used to evaluate the proposed QTL method. In the Figure, (a) represents Ants \& Bees dataset, taken from kaggle (https://www.kaggle.com/datasets/gauravduttakiit/ants-bees), (b) illustrates Road sign detection dataset, taken from kaggle (https://www.kaggle.com/datasets/andrewmvd/road-sign-detection) and (c) represents CIFAR-10 dataset \cite{krizhevsky2009learning}. Each dataset encapsulates distinct challenges and contexts, collectively contributing to a comprehensive assessment of the QTL method's performance. All the datasets are publicly available. This figure is reproduced under terms of the CC-BY license. \cite{khatun2025quantum}. Copyright 2024, Amena Khatun and Muhammad Usman. Advanced Quantum Technologies published by Wiley-VCH GmbH.}
\label{fig:Datasets}
\end{figure*}

\section{EXPERIMENTAL RESULTS}
\subsection{Datasets}

To evaluate the effectiveness and adversarial robustness of the proposed QTL framework, we employed three widely used high-resolution image datasets: Ants \& Bees, CIFAR-10, and Road Sign Detection. Unlike many existing QML/QAML studies that primarily rely on low-resolution datasets such as MNIST (28$\times$28$\times$1) or CIFAR-10 (32$\times$32$\times$3), our evaluation is conducted on higher-dimensional image datasets to demonstrate scalability and real-world applicability.

\textbf{Ants \& Bees} \footnote{\url{https://www.kaggle.com/datasets/gauravduttakiit/ants-bees}} is relatively compact dataset and consists of only 120 labeled images categorised into ants and bees. Additionally, there are 75 validation images for each class. All images have a resolution of 768$\times$512 pixels. Due to the limited size of the dataset, training a deep model from scratch would be inefficient and prone to overfitting. However, our QTL framework capitalises on knowledge transferred from a pre-trained model, allowing it to adapt efficiently and perform well despite the small size datasets.

\textbf{CIFAR-10 (Canadian Institute for Advanced Research)} \cite{krizhevsky2009learning} is a benchmark dataset extensively used in computer vision for evaluating classification algorithms. It contains a total of 60,000 color images across 10 distinct object categories. The training subset comprises 50,000 samples, evenly distributed among the classes, while the test subset consists of 10,000 images. Each image is of dimension 32$\times$32 pixels and includes three color channels (RGB). Although the images are low in resolution, the dataset remains challenging due to intra-class variability and visual complexity across categories.

\textbf{Road Sign Detection Dataset} \footnote{\url{https://www.kaggle.com/datasets/andrewmvd/road-sign-detection}} is a high-resolution dataset contains 877 images, each with a resolution of 1024$\times$1024 pixels. It is curated to support the detection and classification of road signage. The dataset encompasses four primary categories: traffic lights, stop signs, speed limit signs, and pedestrian crossings. We employed this dataset to train and validate our QTL model for traffic sign classification tasks, a critical component in intelligent transportation systems and autonomous vehicle applications. For experimental consistency, the dataset was divided into an 80\% training set and a 20\% testing set.

\subsection{Performance Comparison Between QTL and Classical TL Approaches}

\begin{table}[!t]
%\fontsize{9}{9}\selectfont
\centering
\begin{tabular}{|p{3.0cm}|p{2.8cm}|p{1.5cm}|}
\hline
\textbf{Method} & \textbf{Dataset} & \textbf{Accuracy}\\
\hline\hline
Classical TL &Ants \& Bees &94.70\% \\
Classical without TL  &Ants \& Bees &62.75\% \\
Quantum TL  &Ants \& Bees &96.10\%\\
Quantum without TL  &Ants \& Bees &65.36\% \\
\hline
Classical TL &CIFAR-10 &92.10\% \\
Classical without TL  &CIFAR-10 &72.9\% \\
Quantum TL  &CIFAR-10 &95.80\%\\
Quantum without TL  &CIFAR-10 &51.70\% \\
\hline
Classical TL &Road Sign Detection &92.47\% \\
Classical without TL  &Road Sign Detection &60.22\% \\
Quantum TL  &Road Sign Detection &94.62\%\\
Quantum without TL  &Road Sign Detection &48.39\%
\\
\hline
\end{tabular}
\caption{Comparison of classification accuracy across Classical TL, Classical without TL, Quantum TL, and Quantum without TL methods on Ants \& Bees \cite{paszke2019pytorch}, CIFAR-10 \cite{krizhevsky2009learning}, and Road Sign Detection \cite{mogelmose2012vision} datasets. This table is reproduced under terms of the CC-BY license. \cite{khatun2025quantum}. Copyright 2024, Amena Khatun and Muhammad Usman. Advanced Quantum Technologies published by Wiley-VCH GmbH.}
\label{tab:TL_results}
\end{table}

The results of our QTL framework against classical TL baselines across three datasets are reported in Table \ref{tab:TL_results}. For the Ants \& Bees dataset, the QTL model achieves a classification accuracy of 96.1\%, surpassing the classical TL model which attained 94.7\%. Similarly, on the CIFAR-10 and Road Sign Detection datasets, QTL attained accuracies of 95.8\% and 94.62\%, outperforming classical TL models that achieved 92.1\% and 92.47\%, respectively. These results demonstrate consistent improvements of 1.4\%, 3.7\%, and 2.2\% in classification accuracy across the datasets when using the QTL framework. To further validate the importance of TL in both classical and quantum contexts, we also evaluated performance when training the models from scratch, i.e., without TL. The architectural setups used for these evaluations followed the designs outlined in Figure \ref{fig:QTL_Architecture} and Section \ref{without_TL}. Without TL, the quantum model achieved 65.36\% accuracy on Ants \& Bees, representing a 30.74\% drop compared to the QTL counterpart. A similar decline is evident in the CIFAR-10 and Road Sign Detection datasets, where quantum models trained without TL reached 51.7\% and 48.39\% accuracy, respectively, reflecting performance drops of 44.1\% and 46.23\% compare to QTL. This notable degradation underscores the importance of TL in enhancing model performance, particularly in quantum neural networks, by enabling effective feature reuse and faster convergence. These empirical findings reinforce the effectiveness of the proposed QTL framework compared to classical and non-transfer learning quantum models.

\subsection{Adversarial Attack Evaluation and Robustness through Adversarial Training}
\label{AT_results}
\begin{table}[!t]
%\fontsize{9}{9}\selectfont
\centering
\begin{tabular}{|p{2.5cm}|p{1.2cm}|p{1.3cm}|p{1.3cm}|p{2.2cm}||p{2.0cm}|}
\hline
\textbf{Method} & \textbf{Attack Strength} & \textbf{Clean \newline Accuracy} & \textbf{Accuracy under  \newline Attack} & \textbf{Clean \newline Accuracy with \newline adversarial data} & \textbf{Adversarial Training  \newline Accuracy}\\
\hline\hline
Classical TL & 0.1 &94.70\% &50.64\% &92.38\% &64.10\% \\
Classical TL & 0.2 &94.70\% &47.70\% &89.17\% &61.10\% \\
Classical TL & 0.3 &94.70\% &45.79\% &86.93\% &59.10\% \\
Classical without TL & 0.1 &62.75\% &44.87\% &59.48\% &47.64\% \\
Classical without TL & 0.2 &62.75\% &44.50\% &57.54\% &47.01\% \\
Classical without TL & 0.3 &62.75\% &44.23\% &56.44\% &46.79\% \\
\hline
Quantum TL & 0.1 &96.10\% &53.85\% &94.76\% &71.87\% \\
Quantum TL & 0.2 &96.10\% &50.90\% &94.53\% &69.30\% \\
Quantum TL & 0.3 &96.10\% &47.45\% &94.12\% &67.82\% \\
Quantum without TL & 0.1 &65.36\% &45.68\% &49.23\% &48.12\% \\
Quantum without TL & 0.2 &65.36\% &45.12\% &48.74\% &48.0\% \\
Quantum without TL & 0.3 &65.36\% &45.12\% &48.06\% &47.98\% \\
\hline
\end{tabular}
\caption{Simulation results of the proposed QTL approach on Ants \& Bees dataset under adversarial attack with a different attack strength are compared to the accuracy of classical TL under attack. The experiments are also performed on AT to evaluate the robustness of the proposed method. This table is reproduced under terms of the CC-BY license. \cite{khatun2025quantum}. Copyright 2024, Amena Khatun and Muhammad Usman. Advanced Quantum Technologies published by Wiley-VCH GmbH.}
\label{tab:Ants_bees_AT}
\end{table}

\begin{table}[!htbp]
\begin{center}
%\begin{table*}[!t]
%\fontsize{7}{7}\selectfont
\centering
\begin{tabular}{|p{1.6cm}|p{1.5cm}|p{1.5cm}|p{1.2cm}|p{1.55cm}|}
\hline
\textbf{Method} & \textbf{Attack Strength} & \textbf{Clean \newline Accuracy} & \textbf{Accuracy under Attack} 
& \textbf{Adversarial \newline Training \newline Accuracy}\\
\hline\hline
Classical TL & 0.1 &93.7\% &85.4\%  &93.7\% \\
Classical TL & 0.3 &93.7\% &83.3\%  &92.4\% \\
Classical TL & 1.0 &93.7\% &58.7\% &89.5\% \\
\hline
Quantum TL & 0.1 &95.8\% &89.6\% &95.8\% \\
Quantum TL & 0.3 &95.8\% &87.5\%  &93.7\% \\
Quantum TL & 1.0 &95.8\% &60.4\%  &92.1\% \\
\hline
\end{tabular}
\end{center}
\caption{Simulation results of the proposed QTL approach on Road Sign detection dataset under adversarial attack with a different attack strength is compared to the accuracy of classicalTL under attack. The experiments are also performed on AT to evaluate the robustness of the proposed method. This table is reproduced under terms of the CC-BY license. \cite{khatun2025quantum}. Copyright 2024, Amena Khatun and Muhammad Usman. Advanced Quantum Technologies published by Wiley-VCH GmbH.}
\label{tab:road_data}
\end{table}

\begin{table}[!htbp]
\begin{center}
%\begin{table*}[!t]
%\fontsize{7}{7}\selectfont
\centering
\begin{tabular}{|p{1.6cm}|p{1.5cm}|p{1.5cm}|p{1.2cm}|p{1.55cm}|}
\hline
\textbf{Method} & \textbf{Attack Strength} & \textbf{Clean \newline Accuracy} & \textbf{Accuracy under Attack} & \textbf{Adversarial \newline Training \newline Accuracy}\\
\hline\hline
Classical TL & 0.1 &92.1\% &50.1\% &85.8\% \\
Classical TL & 0.3 &92.1\% &50.2\% &84.9\% \\
Classical TL & 1.0 &92.1\% &49.2\% &80.8\% \\
\hline
Quantum TL & 0.1 &95.8\% &50.9\% &86.6\% \\
Quantum TL & 0.3 &95.8\% &50.0\%  &86.2\% \\
Quantum TL & 1.0 &95.8\% &50.0\% &85.0\% \\
\hline
\end{tabular}
\end{center}
\caption{Simulation results of the proposed QTL approach on CIFAR-10 dataset under adversarial attack with a different attack strength is compared to the accuracy of classical TL under attack. The experiments are also performed on AT to evaluate the robustness of the proposed method. This table is reproduced under terms of the CC-BY license. \cite{khatun2025quantum}. Copyright 2024, Amena Khatun and Muhammad Usman. Advanced Quantum Technologies published by Wiley-VCH GmbH.}
\label{tab:cifar-10}
\end{table}

To evaluate the robustness of our proposed QTL framework against adversarial attack, we subjected the trained model to adversarial perturbations generated using the FGSM method, as described in Section \ref{adversarial_attack}. The results for Ants \& Bees dataset is reported Table \ref{tab:Ants_bees_AT}. We investigated the impact of adversarial noise at different perturbation magnitudes ($\epsilon$), specifically at $\epsilon = 0.1$, $0.2$, and $0.3$, across four settings: Classical TL, Classical without TL, Quantum TL, and Quantum without TL. For the QTL model, we observe a sharp drop in performance from a clean accuracy of 96.1\% to 53.85\% when subjected to an adversarial strength of $\epsilon = 0.1$, corresponding to a 42.25\% decrease. Increasing the attack strength to $\epsilon = 0.2$ causes a further degradation of 2.95\%. The deterioration in classification accuracy is a clear indication that the quantum TL model is vulnerable against adversarial attack as the classical TL method. 

To address this vulnerability, we introduced adversarial training (detailed in Section \ref{adversarial_training}) into the QTL pipeline. When the QTL model is adversarially trained at perturbation levels of $\epsilon = 0.1$, $0.2$, and $0.3$, the model achieves significantly higher accuracies of 71.87\%, 69.30\%, and 67.82\%, respectively. These results represent notable gains of 18.02\%, 18.4\%, and 20.35\% compared to their non-robust QTL counterparts under attack. We also report results for QNN without TL, reinforcing the insight that transfer learning greatly enhances both clean and adversarial robustness performance. Furthermore, we observed that classical models without TL exhibit substantial performance degradation under adversarial influence, underscoring the critical role of TL in building robust representations. Our findings are aligned with prior literature such as \cite{ijcai2021p591}, which also demonstrated significant susceptibility of classical CNNs under adversarial attack.

\begin{figure}
\includegraphics[width=0.9\linewidth]{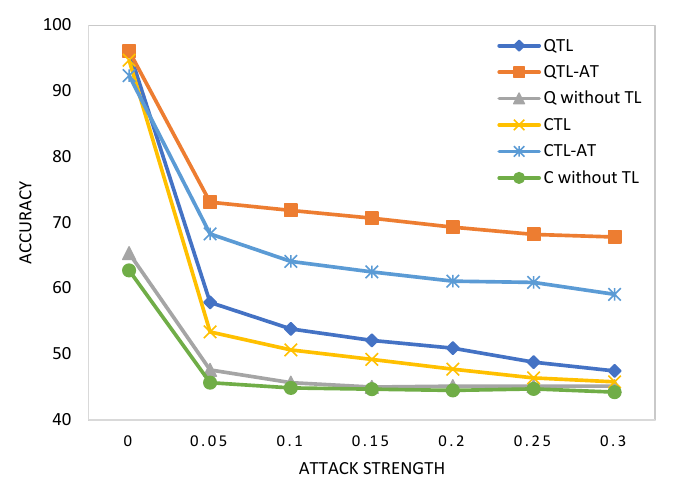}
   \caption{The plot of accuracy versus attack strength on Ants \& Bees for QTL (quantum transfer learning), QTL-AT (quantum transfer learning with adversarial training), Q without TL (quantum without transfer learning), CTL (classical transfer learning), CTL-AT (classical transfer learning with adversarial training), C without TL (classical without transfer learning). This figure is reproduced under terms of the CC-BY license. \cite{khatun2025quantum}. Copyright 2024, Amena Khatun and Muhammad Usman. Advanced Quantum Technologies published by Wiley-VCH GmbH.}
\label{fig:graph}
\end{figure}

The effectiveness of AT was similarly confirmed on the Road Sign Detection and CIFAR-10 datasets, with detailed results reported in Tables \ref{tab:road_data} and \ref{tab:cifar-10}. For Road Sign Detection, applying an adversarial attack of strength $\epsilon = 0.1$ caused a drop in performance from 95.8\% to 89.6\%. Interestingly, the performance only declined by an additional 2.1\% when $\epsilon$ increased to 0.3. Hence, we increase the attack strength to 1.0 and notice a significant reduction of model performance to 60.4\%. On the CIFAR-10 dataset, the QTL model's accuracy decreased marginally from 50.9\% to 50.0\% under adversarial attacks of increasing intensity ($\epsilon = 0.1$, $0.3$, and $1.0$). This suggests a relative robustness on CIFAR-10 dataset, although performance still benefits from AT. Across both datasets, applying AT substantially mitigated performance loss, with accuracy levels approaching the clean baseline for the Road Sign Detection dataset in particular. Figure \ref{fig:graph} visualises model accuracy as a function of adversarial strength $\epsilon$ for multiple settings: QTL, QTL-AT (QTL with adversarial training), quantum without TL, Classical TL (CTL), CTL-AT (CTL with adversarial training), and classical without TL. The QTL-AT curve consistently outperforms all other approaches across perturbation levels, clearly demonstrating the superior adversarial resilience of our proposed method.

\section{Analysis and Interpretation of Results}
The results presented in this work demonstrate the effectiveness of the proposed QTL framework, which combines classical and quantum ML techniques for image classification. Across the three benchmark datasets, Ants \& Bees, CIFAR-10, and Road Sign Detection, the QTL model consistently outperformed both classical TL and QNNs trained without TL. One of the main advantages of the QTL approach is its ability to address the limitations of current quantum resources, such as limited number of qubits, while still handling high-dimensional input data. This is achieved by using a pre-trained classical CNN (ResNet-18) as a feature extractor. The ResNet-18 model extracts relevant and compact representations from raw images, which are then passed to a QVC for further processing. This knowledge transfer from a large-scale classical CNN to a small-scale QVC-based QNN leads the proposed QTL to achieve gain in performance. The results clearly highlight the benefits of this hybrid approach. On the Ants \& Bees dataset, the QTL model achieved 96.1\% accuracy, which is 1.4\% higher than the classical TL model and 30.74\% higher than QNN without TL. Similar performance improvements were observed for CIFAR-10 and Road Sign Detection datasets, confirming that the QTL framework can generalise well across different tasks. In addition to evaluating performance under clean data conditions, we also tested the robustness of the QTL model against adversarial attacks. As described in Section \ref{AT_results}, adversarial attacks led to a significant drop in model accuracy. However, when AT was incorporated within the proposed framework, the performance of the QTL model improved notably, demonstrating that AT is effective in increasing the model's resistance to perturbations. This improvement was observed across all datasets, highlighting the importance of integrating robustness techniques into QTL framework.

Quantum computing systems are known to produce computational patterns that are counter-intuitive and believed to be beyond the efficient reach of classical hardware \cite{biamonte2017quantum}. It has been suggested that by exploiting the quantum superposition and entanglement, quantum models are capable of processing and representing information in fundamentally different ways than their classical counterparts. Recent work has also shown that QML models, including those explored in this study, exhibit a higher degree of robustness against adversarial perturbations \cite{west2023benchmarking}, a property that naturally arises from the inherent characteristics of quantum mechanics. We believe that combining quantum and classical models via TL captures the advantages of both domains by leveraging the representational power of quantum systems while utilising the learning efficiency of classical models. The performance gains observed in our experiments suggest that hybrid quantum-classical models such as QTL are capable of outperforming purely classical approaches, particularly in terms of both accuracy and robustness. Since classical systems lack access to quantum phenomena, , it is not possible to train classical systems to match the performance of quantum models. Our results indicate that QTL can act as a bridge between today's limited quantum hardware resources and the demands of high-dimensional learning tasks, even when limited training data is available. In future work, we aim to deploy and benchmark our QTL models on real quantum processors to test their practical performance in real-world applications, particularly when fault-tolerant quantum hardware becomes accessible in the coming years.

\section{Conclusion}
In this work, we introduced a comprehensive QTL framework that integrates the capabilities of quantum computing and classical ML into a unified architecture. Extensive simulations conducted on diverse datasets, including Ants \& Bees, CIFAR-10, and Road Sign Detection demonstrated the superiority of the proposed QTL approach over traditional classical TL methods. The results consistently showed that QTL achieved higher classification accuracy across all datasets, highlighting its effectiveness even in scenarios with limited training data. Additionally, we explored the susceptibility of QTL model to adversarial attacks. By applying adversarial perturbations to the input data, we exposed potential security vulnerabilities in quantum-enhanced models. To address this, we incorporated AT into the QTL framework, which significantly improved the resilience and overall classification performance under attack scenarios. The results from our comparative analysis underscore the potential of QTL in advancing QML applications, particularly in improving accuracy and robustness. Our simulation results and extensive comparison of QTL with its classical counterparts open up avenues for future research in quantum-enhanced ML.

\section*{Acknowledgments}
The authors acknowledge the use of CSIRO HPC (High-Performance Computing) and NCI's Gadi supercomputer for conducting the experiments. A.K. and M.U. also acknowledge CSIRO's Quantum Technologies Future Science Platform for providing the opportunity to work on quantum machine learning.

\bibliographystyle{unsrt}

\bibliography{references}
\end{document}